\documentclass[12pt]{article}
\def \efi{Enrico Fermi Institute Report No.~}
\def \ka{K_1^0}
\def \kb{K_2^0}
\def \ok{\overline{K}^0}
\begin{document}
\baselineskip 22pt
\centerline{\bf CP SYMMETRY VIOLATION
\footnote{\efi 01-40, hep-ph/0109240, to be published in {\it Macmillan
Encylopedia of Physics, Supplement:  Elementary Particle Physics},
edited by John S. Rigden, Jonathan Bagger, and Roger H. Stuewer
(Macmillan Reference USA, New York, 2002).}}
\centerline{Jonathan L. Rosner}
\medskip

The symmetry known as CP is a fundamental relation between matter and
antimatter.  The discovery of its violation by Christenson, Cronin, Fitch,
and Turlay (1964) has given us important insights into the structure of
particle interactions and into why the Universe appears to contain more matter
than antimatter. 

In 1928, Paul Dirac predicted that every particle has a
corresponding {\it antiparticle}.  If the particle has {\it quantum numbers}
(intrinsic properties),
such as electric charge, the antiparticle will have opposite quantum numbers.
Thus, an electron, with charge $-|e|$, has as its antiparticle a {\it positron},
with charge $+|e|$ and the same mass and spin.  Some neutral particles,
such as the {\it photon}, the quantum of radiation, are their own
antiparticles.  Others, like the neutron, have distinct antiparticles; the
neutron carries a quantum number known as {\it baryon number} B = 1,
and the antineutron has B = --1.  (The prefix {\it bary-} is Greek for
{\it heavy}.)  The operation of {\it charge reversal}, or
C, carries a particle into its antiparticle.

Many laws of physics are {\it invariant} under the C operation; that is,
they do not change their form, and, consequently, one cannot tell whether one
lives in a world made of matter or one made of antimatter.  Many equations are
also invariant under two other important symmetries:  space reflection,
or {\it parity}, denoted by P, which reverses the direction of all spatial
coordinates, and {\it time reversal}, denoted by T, which reverses the arrow
of time.  By observing systems governed by these equations, we cannot tell
whether our world is reflected in a mirror or in which direction its clock is
running.  Maxwell's equations of electromagnetism and the equations of
classical mechanics, for example, are invariant separately under P and T.

Originally it was thought that {\it all} elementary particle interactions were
unchanged by C, P, and T individually.  In 1957, however, it was
discovered that a certain class known as the {\it weak interactions}
(for example, those governing the decay of the neutron) were
not invariant under P or C.  However, they appeared to be invariant under
the product CP and also under T.  (Invariance under the product CPT is
a very general feature of elementary particle theories.)  Thus, it was thought
that one could not distinguish between our world and a mirror-reflected world
made of antimatter, or our world and one in which clocks ran backward.

Murray Gell-Mann and Abraham Pais (1955) used an argument based on C
invariance (recast in 1957 in terms of CP invariance) to discuss the
production and decay of a particle known as the {\it neutral K meson},
or $K^0$.  This particle, according to a theory by Gell-Mann and Kazuo
Nishijima, carried a quantum number called {\it strangeness}, with S($K^0$)
= +1, and so there should exist a neutral anti-K meson, called $\ok$, 
with S($\ok$) = --1.  The theory demanded that strangeness be conserved in
K-meson production but violated in its decay.  Both the $K^0$ and the $\ok$
should be able to decay to a pair of $\pi$ mesons (e.g., $\pi^+ \pi^-$).
How, then, would one tell them apart?

Gell-Mann and Pais solved this problem by applying a basic idea of quantum
mechanics:  The particle decaying to $\pi^+ \pi^-$ would have to have the
same behavior under C (in 1957, under CP) as the final $\pi^+ \pi^-$
combination, which has CP = +1.  (That is, its quantum-mechanical state is
taken into itself under the CP operation.)
A quantum-mechanical combination of $K^0$
and $\ok$ with this property was called $\ka$.  There should then exist
another combination of $K^0$ and $\ok$ with CP = --1 (i.e., its
quantum-mechanical state is changed in sign under the CP operation).
This particle was called $\kb$.
(The subscripts 1 and 2 were used simply to distinguish the two particles
from one another.)  The $\kb$ would
be forbidden by CP invariance from decaying to $\pi \pi$ and thus, being
required to decay to three-body final states, would be much longer-lived.
This predicted particle was discovered in 1956.

Christenson, Cronin, Fitch, and Turlay performed their historic experiment in
the early 1960s at Brookhaven National Laboratory to see if the long-lived
neutral K meson could occasionally decay to $\pi^+ \pi^-$.  They found that
indeed it did, but only once every 500 decays.  For this discovery Cronin and
Fitch were awarded the 1980 Nobel Prize in Physics.

The short-lived neutral
K meson was renamed $K_S$ and the long-lived one $K_L$.  The $K_L$ lives
nearly 600 times as long as the $K_S$.  The discovery of its decay to $\pi^+
\pi^-$ was the first evidence for violation of CP symmetry.  The $K_S$ is
mainly CP-even, while the $K_L$ is mainly CP-odd.  Within any of the current
interaction theories, which conserve the product CPT, the violation of CP
invariance then also implies T-invariance violation.

Shortly after CP violation was detected, Andrei Sakharov (1967) proposed that
it was a key ingredient in understanding why the Universe is composed of more
matter than antimatter.  Another ingredient in his theory was the need for
baryon number (the quantum number possessed by neutrons and protons) to be
violated, implying that the proton should not live forever.  The search for
proton decay is an ongoing topic of current experiments.

CP violation can also occur in {\it quantum chromodynamics} (QCD), the theory
of the strong interactions, through solutions which violate both P and T.
However, this form of CP violation appears to be extremely feeble, less than a
part in ten billion; otherwise it would have contributed to detectable effects
such as {\it electric dipole moments} of neutrons.  It is not
known why this form of CP violation is so weak; proposed solutions to the
puzzle include the existence of a light neutral particle known as the
{\it axion}.

The leading theory of CP violation was posed by Makoto Kobayashi and Toshihide
Maskawa (1973).  Weak coupling constants of {\it quarks} (the subunits
of matter first postulated in 1964 by Gell-Mann and George Zweig) can have
both real and imaginary parts.  These complex phases lead not only
to the observed magnitude of CP violation discovered by Christenson {\it et
al.}, but also to small differences in the ratios of $K_S$ and $K_L$ decays to
pairs of charged and neutral $\pi$ mesons (confirmed by experiments at CERN and
Fermilab), and to differences in decays of neutral $B$ mesons and their
antiparticles.  Experiments at the Stanford Linear Accelerator Center (SLAC)
using the BaBar detector (named after the character in the children's book) and
at the National Laboratory for High Energy Physics in Japan (KEK) using the
Belle detector have recently reported convincing evidence for this last effect
(Aubert {\it et al.}, 2001; Abe {\it et al.}).  At a deeper level, however,
both the origin of the matter-antimatter asymmetry of the Universe discussed
by Sakharov and the source of the complex phases of Kobayashi and Maskawa
remain a mystery, perhaps stemming from some common source.
\newpage

\baselineskip=20pt
\centerline{\bf BIBLIOGRAPHY}

\leftline{\underline{Books}:}

\noindent
BIGI, I. I. and SANDA, A. I. {\it CP Violation}.  Cambridge: Cambridge
University Press, 2000.
\medskip

\noindent
BRANCO, G. C., LAVOURA, L, and SILVA, J. P. {\it CP Violation}.  Oxford:
Clarendon Press, 1999.
\bigskip

\noindent
\leftline{\underline{Journal articles}:}
\medskip

\noindent
ABE, K. {\it et al.} ``Observation of Large CP Violation in the Neutral B
Meson System.'' {\it Physical Review Letters} 87 (2001) 091802.
\medskip

\noindent
AUBERT, B. {\it et al.} ``Observation of CP Violation in the $B^0$ Meson
System.'' {\it Physical Review Letters} 87 (2001) 091801.
\medskip

\noindent
CHRISTENSON, J. H., CRONIN, J. W., VITCH, V. L.; and TURLAY, R. E.  ``Evidence
for the $2 \pi$ Decay of the $\kb$ Meson.''  {\it Physical Review Letters}
13 (1964) 138--140.
\medskip

\noindent
CRONIN, J. W.  ``CP Symmetry Violation:  The Search for its Origin.''
{\it Reviews of Modern Physics} 53 (1981) 373--383.
\medskip

\noindent
FITCH, V. L.  ``The Discovery of Charge Conjugation Parity Asymmetry.''
{\it Reviews of Modern Physics} 53 (1981) 367--371.
\medskip

\noindent
GELL-MANN, M. and PAIS, A.  ``Behavior of Neutral Particles Under Charge
Conjugation.''  {\it Physical Review} 97 (1955) 1387--1389.
\medskip

\noindent
KOBAYASHI, M. and MASKAWA, T.  ``CP Violation in the Renormalizable Theory
of Weak Interaction.'' {\it Progress of Theoretical Physics} 49 (1973)
652--657.
\medskip

\noindent
SAKHAROV, A. D.  ``Violation of CP Invariance, C Asymmetry, and Baryon
Asymmetry of the Universe.''  {\it Soviet Physics -- JETP Letters} 5 (1967)
24--27.

\end{document}